\documentclass[aps,prb,preprint,groupedaddress,showpacs,amsmath,amssymb]{revtex4}

\usepackage{graphicx}
\usepackage{dcolumn}
\usepackage{bm}

\begin{document}

\title{Pressure induced superconductivity in CaFe$_2$As$_2$.}

\author{Milton S. Torikachvili}
\affiliation{Department of Physics, San Diego State University, San Diego, CA 92182-1233}
\author{Sergey L. Bud'ko}
\author{Ni Ni}
\author{Paul C. Canfield}
\affiliation{Ames Laboratory US DOE and Department of Physics and Astronomy, Iowa State University, Ames, Iowa 50011}

\date{\today}

\begin{abstract}

CaFe$_2$As$_2$ has been found to be exceptionally sensitive to the application of hydrostatic pressure and superconductivity has been found to exist in a narrow pressure region that appears to be at the interface between two different phase transitions.  The pressure - temperature ($P - T$) phase diagram of CaFe$_2$As$_2$ reveals that this stoichiometric, highly ordered, compound can be easily tuned to reveal all the salient features associated with FeAs-based superconductivity without introducing any disorder.  Whereas at ambient pressure CaFe$_2$As$_2$ does not superconduct for $T > 1.8$ K and manifests a first order structural phase transition near $T \approx 170$ K, the application of $\sim 5$ kbar hydrostatic pressure fully suppresses the resistive signature of the structural phase transition and instead superconductivity is detected for $T < 12$ K.  For $P \ge 5.5$ kbar a different transition is detected, one associated with a clear reduction in resistivity and for $P > 8.6$ kbar superconductivity is no longer detected.  This higher pressure transition temperature increases rapidly with increasing pressure, exceeding 300 K by $P \sim 17$ kbar.  The low temperature, superconducting dome is centered around 5 kbar, extending down to 2.3 kbar and up to 8.6 kbar.  This superconducting phase appears to exist when the low pressure transition is suppressed sufficiently, but before the high pressure transition has reduced the resistivity, and possibly the associated fluctuations, too dramatically.

\end{abstract}


\maketitle

Superconductivity has been stabilized in iron arsenide based compounds by fluorine doping, RFeAsO$_{1-x}$F$_x$, \cite{1} oxygen depletion, RFeAsO$_{1-x}$, \cite{2} and potassium doping, Ba$_{1-x}$K$_x$Fe$_2$As$_2$ and Sr$_{1-x}$K$_x$Fe$_2$As$_2$ \cite{3,4} with transition temperatures ($T_c$)  in excess of 50 K for F-doped / O-depleted members of the RFeAsO series and $T_c$ values approaching 40 K for K-doped AFe$_2$As$_2$   (A = Ba, Sr).  The facts that (i) such high transition temperatures have been found in two distinct families of FeAs-based compounds, (ii) these compounds all manifest a structural phase transition that is suppressed by doping, and (iii) doping is necessary to stabilize superconductivity, have given rise to a feeling that these FeAs compounds need to be finely tuned so as to create this intriguing, superconducting state.

The recent discovery of CaFe$_2$As$_2$, a previously unknown member of the $I4/mmm$, AFe$_2$As$_2$ family \cite{ger} has expanded the isostructural series of alkali earth iron arsenides to three:  A = Ca, Sr, Ba. \cite{5,6,7} The structural phase transition from the high temperature, tetragonal phase to the low temperature orthorhombic phase is unambiguously first order in CaFe$_2$As$_2$ and occurs near 170 K with a 2 K hysteresis width. \cite{5} This transition is accompanied by a first order transition to a low temperature commensurate antiferromagnetic state. \cite{ngol} In addition, it has been found that for sodium doping, Ca$_{1-x}$Na$_x$Fe$_2$As$_2$, the structural phase transition is suppressed and superconductivity can be established with $T_c$ values close to 20 K. \cite{6}

Whereas chemical substitution is a convenient method for changing the properties of a compound, it inevitably changes many of the physical parameters in a multitude of uncontrollable ways; there can be changes in the $a$- and $c$- axes of the unit cell, the volume, the band filling, the degree of disorder, etc.  The existing data set on superconductivity in iron arsenide compounds is intriguing, compelling, and exceedingly complex due to the many changes brought on by doping.  There is a growing need to find a system that can be tuned in a more systematic fashion, but that still manifests the salient physics.  In this letter we establish pure CaFe$_2$As$_2$ as just such a system.  The application of very modest hydrostatic pressures ($P < 5$ kbar) suppresses the high temperature tetragonal to orthorhombic phase transition.  For higher pressures ($P > 5.5$ kbar) a second phase transition, with a different resistive signature, is stabilized and increases rapidly with increasing pressure, with its transition temperature exceeding 300 K by 17 kbar.  Nestled between these two, very pressure sensitive transitions is a dome like region of superconductivity with a maximum $T_c$ value of $\sim 12$ K centered close to 5 kbar.  CaFe$_2$As$_2$, then, clearly shows that (i) superconductivity can be stabilized without the complications associated with doping and (ii) superconductivity appears at the interface between the tetragonal to orthorhombic, structural (and antiferromagnetic) phase transition and a second phase transition of an as of yet unknown nature.  As a result, CaFe$_2$As$_2$ may hold the key to understanding the nature and mechanism of superconductivity associated with the whole set of iron arsenide compounds.\\

Single crystals of CaFe$_2$As$_2$ were grown out of a Sn flux, using conventional high temperature solution growth techniques as discussed in Refs. [\onlinecite{5,8}]. The temperature dependence of the in plane resistivity was measured for various hydrostatic pressures below 20 kbar.  Pressure was generated in a Teflon cup filled with Fluorinert FC-75 which was inserted into a 22 mm outer diameter, non-magnetic, piston-cylinder-type, Be-Cu pressure cell with a core made of NiCrAl (40 KhNYu-VI) alloy. The pressure was determined at low temperature by monitoring the shift in the superconducting transition temperature of pure lead \cite{eil81a}. Low temperature pressure values will be used throughout the text.  Errors associated with the determination of the low temperature pressure are $\sim \pm 0.5$ kbar and based on our experience with this cell, the higher temperature transitions will have potential shifts in pressure on the order of $\sim 1$ kbar.  This uncertainty in pressure does not significantly affect any of our conclusions.

The temperature and magnetic field environment for the pressure cell was provided by a Quantum Design Physical Property Measurement System (PPMS-9) instrument. An additional Cernox sensor, attached to the body of the cell, served to determine the temperature of the sample for these measurements. The data presented were taken on cooling. The cooling rate was below 0.5 K/min, the temperature lag between the Cernox on the body of the cell and the system thermometer was $< 0.5$ K at high temperatures and 0.1 K or less below $\sim 70$ K. Below $\sim 10$ K the resistivity was measured in a 250 Oe field so as to suppress the superconductivity of traces of elemental Sn (residual flux).\\

\begin{figure}
\begin{center}
\includegraphics[angle=0,width=80mm]{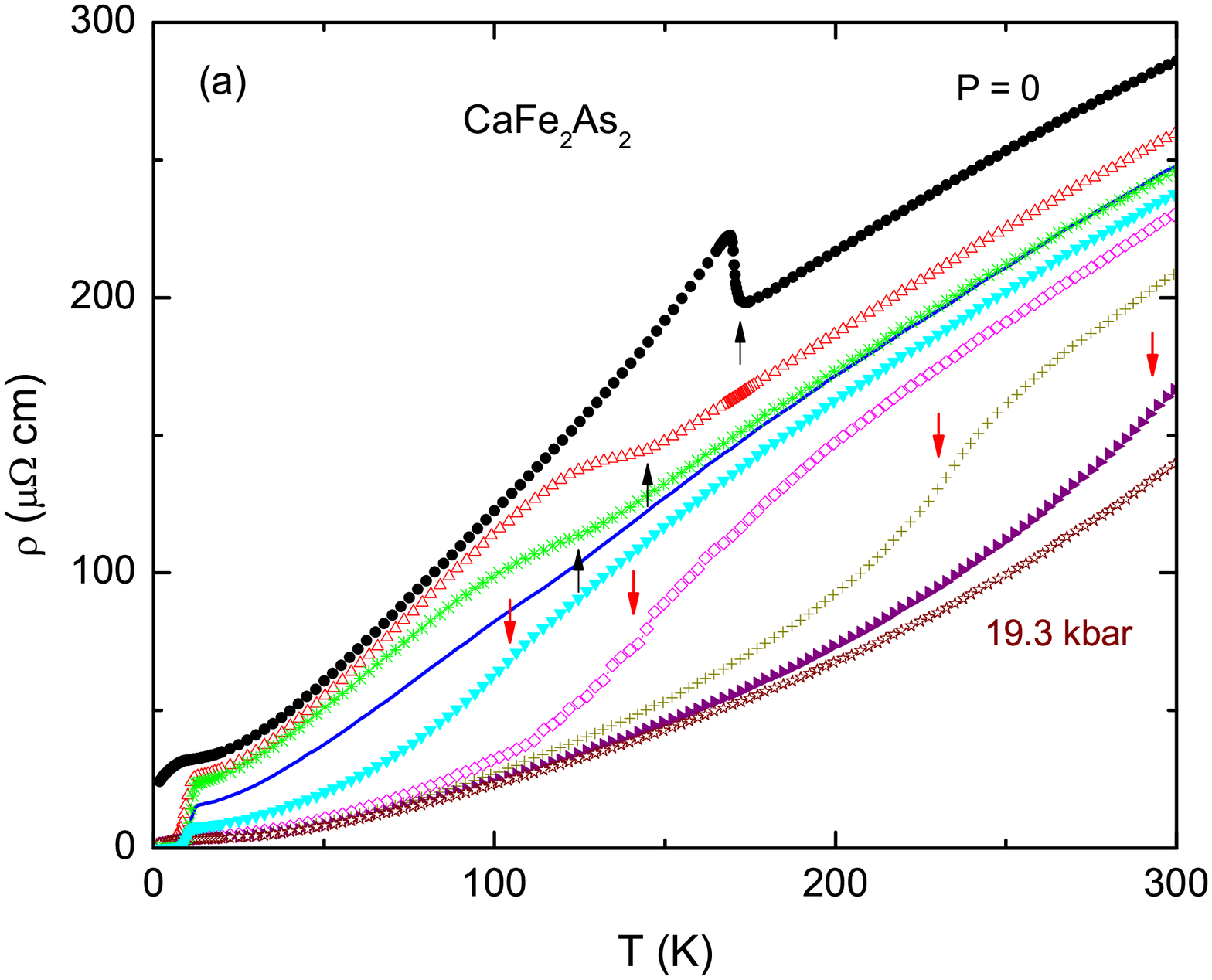}
\includegraphics[angle=0,width=80mm]{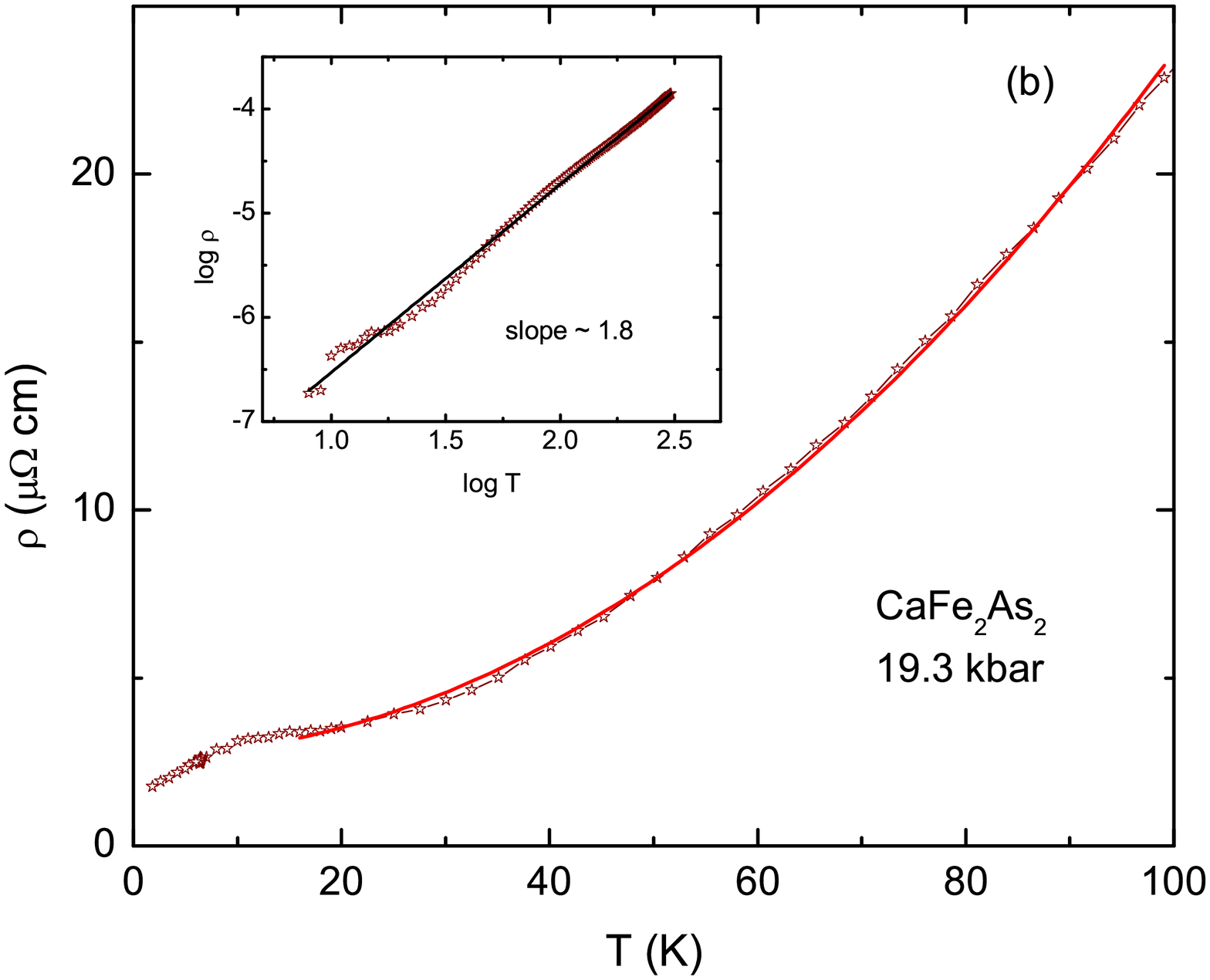}
\includegraphics[angle=0,width=80mm]{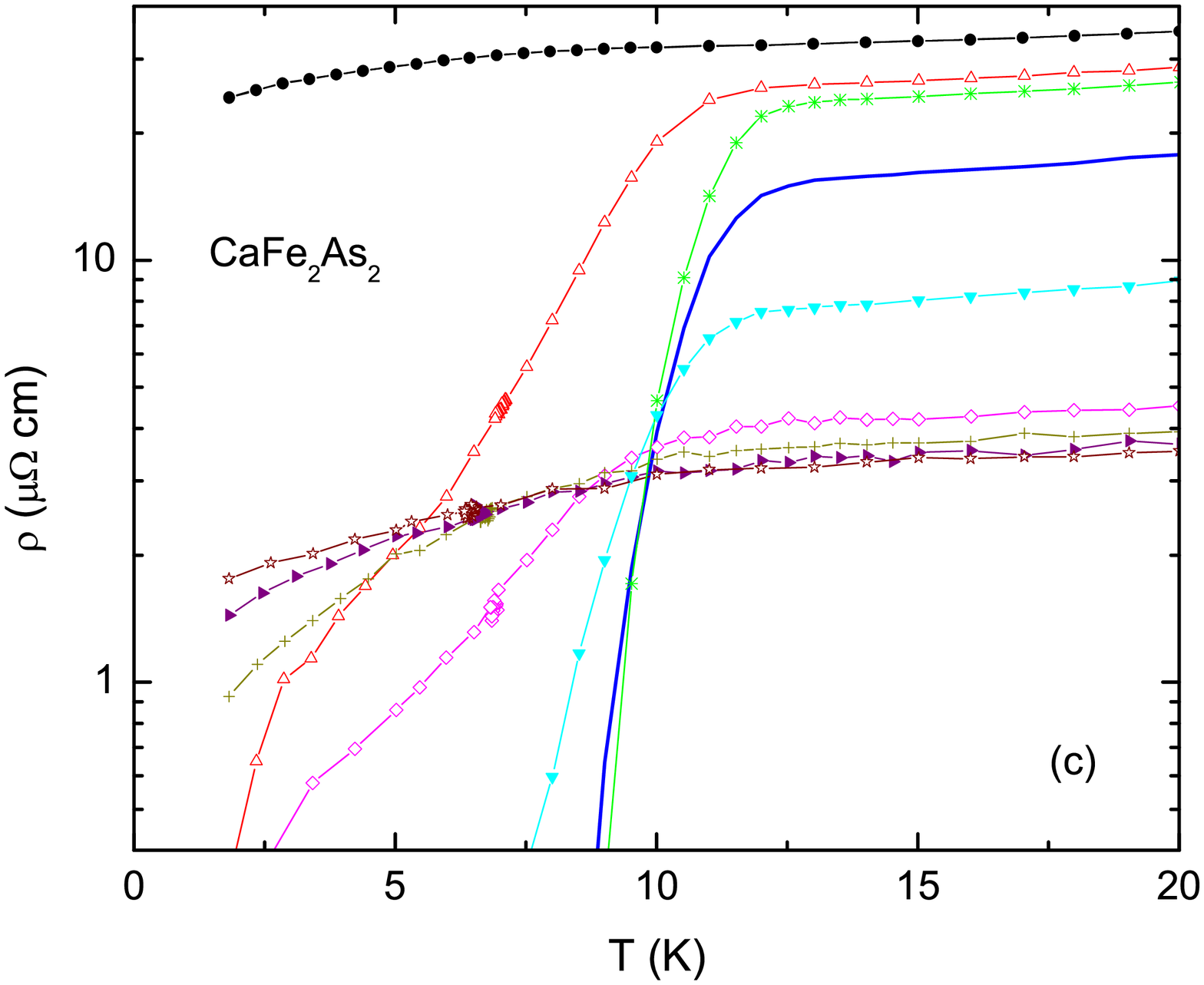}
\end{center}
\caption{(Color online) (a) The in-plane, electrical resistivity of CaFe$_2$As$_2$ as a function of temperature for values of the low temperature pressure,  $P = 0$, 2.3, 3.5, 5.1, 5.5, 8.6, 12.7, 16.8 and 19.3 kbar.  The downward and upward pointing arrows indicate the location of the upper transitions temperatures (see text). (b) The in-plane, electrical resistivity of CaFe$_2$As$_2$ as a function of temperature for $P = 19.3$ kbar (symbols) and fit of data for 15 K $< T <$ 100 K to the form $\rho = \rho_0 + AT^2$ (solid line, the fit results are $\rho_0 = 2.6832 \mu\Omega$ cm, $A = 2.09487 \times 10^{-3} \mu\Omega$ cm/K$^2$).  Inset: {\it log-log} plot of $\rho(T)-\rho_0$ data shown for 10 K $< T <$ 300 K. (c) Low temperature expansion of data shown in panel (a) shown on a {\it semi-log} plot so as to clearly present details for all applied pressures despite a dramatic drop in the residual resistivity at higher pressures.  Symbols are the same as those used in panel (a).} \label{F1}
\end{figure}

Figure \ref{F1} presents the temperature dependent, basal plane, electrical resistivity of single crystalline CaFe$_2$As$_2$ for applied pressures ranging from near atmospheric to approaching 20 kbar.  These data are remarkable in that there are three, very pressure dependent features that can be seen, all below 20 kbar.  The first feature is the conspicuous, discontinuous jump in electrical resistivity near 170 K seen in the ambient pressure resistivity.  This feature has been clearly associated with a first order phase transition from the high temperature, tetragonal phase to a low temperature, orthorhombic (antiferromagnetic) phase.  As pressure is increased to 2.3 kbar this feature drops to $\sim 145$ K and broadens but remains first order with the same $\sim 2$ K thermal hysteresis.  For the next pressure, 3.5 kbar, this feature drops further, to $\sim 130$ K.  It is important to note that the resistivity of CaFe$_2$As$_2$, well below this transition (say between 20 and 50 K) is similar for each of these three data sets, suggesting that the nature of the low temperature state is similar.

The second conspicuous feature is the higher pressure transition that is most clearly seen in the 12.7 kbar data.  For this pressure there is a dramatic drop in resistivity that starts below $\sim 250$ K.  This feature can also be clearly seen for several pressures below 12.7 kbar and may also be present in the $P = 16.8$ kbar data set as well. This loss of resistivity transition is extremely hysteretic, manifesting up to $\sim 30$ K offsets between warming and cooling scans. For $P = 8.6$ kbar and higher, the low temperature (i.e. sufficiently below the phase transition), temperature dependence of the resistivity is identical, again consistent with the low temperature state of CaFe$_2$As$_2$ being the same for $P \ge 8.6$ kbar.

The $P = 19.3$ kbar data appears to represent a limiting curve for the high pressure data, with the 5.5 kbar, 8.6 kbar, 12.7 kbar and 16.8 kbar data increasingly falling upon it as the decrease in resistivity transition temperature is increased by increasing pressure.  Figure \ref{F1}(b) presents the 19.3 kbar data as well as a fit to the form of $\rho = \rho_0 + AT^2$ for data between 15 and 100 K.  Below 15 K the data manifest the shallow drop in resistivity seen in the ambient pressure as well as the $P > 8.6$ kbar data (discussed below), for this reason the $T < 10$ K data are excluded from this power law parametrization.  The fit shown as a solid line in Fig. \ref{F1}(b) is rather good.  The inset to Fig. \ref{F1}(b) is a {\it log-log} plot of the 19.3 kbar data (minus the $\rho_0$ term inferred above).  These data show that there is not perfect, power-law behavior, but that there is a fair fit to these data between 10 and 300 K for $\rho = \rho_0 + AT^n$ with $n = 1.8$.

The third feature is most clearly seen in Figure \ref{F1}(c).  For pressures near 5 kbar there is a clear and complete superconducting transition.  The superconducting groundstate is absent at ambient pressure and is also absent for pressures greater than 8.6 kbar.  Whereas this transition is complete (i.e. we measure zero resistivity at low temperature) for $P = 3.5,~5.1$, and 5.5 kbar, for $P = 2.3$ and 8.6 kbar there is a sharp drop in resistivity that is highly suggestive of a superconducting phase transition, but the resistivty does not go completely to zero, even by 2.0 K.  It is worth noting that the ambient pressure data, as well as the data for $P > 8.6$ kbar, manifest a broad down turn in resistivity for $T < 10$ K.  The cause of this broad, gradual and only partial reduction of resistivity is, as of yet, unknown.

\begin{figure}
\begin{center}
\includegraphics[angle=0,width=90mm]{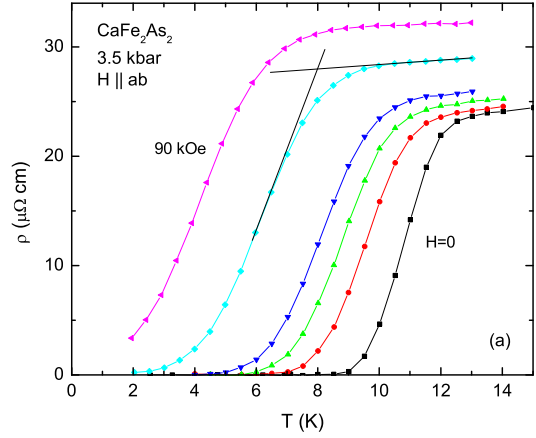}
\includegraphics[angle=0,width=90mm]{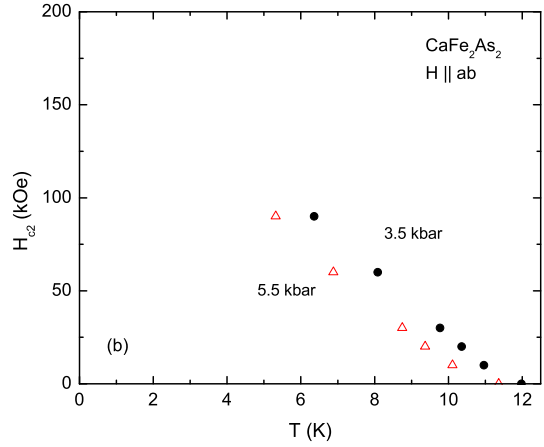}
\end{center}
\caption{(Color online) (a) Low temperature, in plane, electrical resistivity of single crystalline CaFe$_2$As$_2$ for applied magnetic field perpendicular to the crystallographic $c$-axis, for $P = 3.5$ kbar.  Data sets for applied fields of 0, 10, 20, 30, 60 and 90 kOe are shown with an "onset criterion" for $T_c$ shown for the 60 kOe data set (see text). (b) $H_{c2}(T)$ of CaFe$_2$As$_2$ for applied magnetic field perpendicular to the crystallographic $c$-axis for $P = 3.5$ kbar (filled circles) and 5.5 kbar (open triangles).}\label{F2}
\end{figure}

In order to more fully characterize the superconducting state, resistance data were collected for magnetic fields, $H \le 90$ kOe, applied perpendicular to the crystallographic $c$-axis when $P = 3.5$ kbar.  As can be seen in Figure \ref{F2}(a), there is a monotonic suppression of the superconducting transition, with modest broadening (from 2.5 K width in zero field to 4.0 K width in an applied field of 60 kOe).  Using a resistive onset criterion (shown for the 60 kOe data in Fig. \ref{F2}(a)), $H_{c2}(T)$ data can be inferred and are plotted in Fig. \ref{F2}(b).  These data are consistent with a low temperature ($T = 0$) $H_{c2}$ value between 190 and 120 kOe, depending on the form of extrapolation or model used.  Similar data were collected for $P = 5.5$ kbar (not shown) and $H_{c2}(T)$ for this pressure are also presented in Fig. \ref{F2}(b).  There is a slight downward shift in $H_{c2}(T)$ curve for the 5.5 kbar data set, consistent with a slight decrease in the zero field $T_c$ value.  Although these data are not conclusive proof that the low temperature state of CaFe$_2$As$_2$ for $P = 3.5$ and 5.5 kbar of hydrostatic pressure is superconducting, they are extremely compelling.  Further measurements of either magnetization or specific heat under pressure will be needed to determine the superconducting fraction of the sample at these pressures.\\

\begin{figure}
\begin{center}
\includegraphics[angle=0,width=120mm]{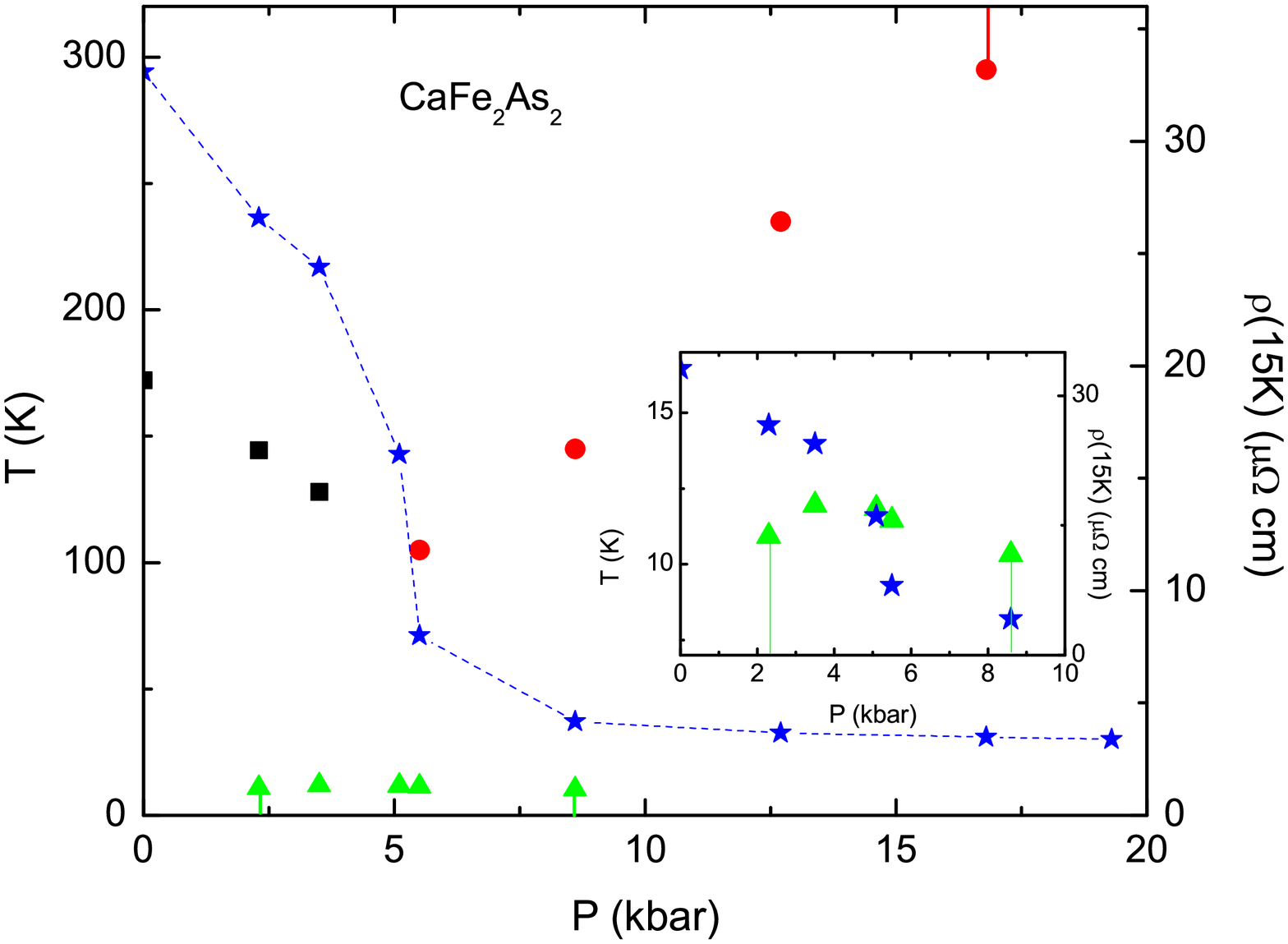}
\end{center}
\caption{(Color online) Pressure - temperature phase diagram of CaFe$_2$As$_2$.  Filled squares represent lower pressure transitions from high temperature tetragonal phase to the lower temperature orthorhombic (antiferromagnetic) phase.  Filled circles represent the higher pressure phase transition that is evidenced by a marked loss of low temperature resistivity.  Filled triangles represent the lower temperature transition to the superconducting state.  The filled stars are the 15 K resistivity values (plotted against the right hand axis).  The inset shows more clearly how the superconducting dome is centered around the sharp loss of low temperature resistivity near 5 kbar.}\label{F3}
\end{figure}

The data in Fig. \ref{F1} can be summarized in the $P - T$ phase diagram shown in Fig. \ref{F3}.  The criteria used for determining the transition temperature for each phase transition do not change the qualitative nature of this figure, but they do affect the figure quantitatively.  For this reason it is important to clearly outline each criterion.  For the low-pressure, structural phase transition we use the high temperature, break in slope, indicated by the upward pointing arrows in Fig \ref{F1}(a).  For the high pressure, drop in resistivity transition, we use the temperature at which $d \rho/d T$ is maximum (maximum slope) to define the transition temperature, indicated in Fig. \ref{F1}(a) by downward pointing arrows.  For the superconducting phase transition, we use the onset criterion outlined in the discussion of $H_{c2}$ and Fig. \ref{F2}.

The two, higher temperature, transitions are very pressure sensitive, with differing signs of $dT_{crit}/dP$.  The low pressure, structural phase transition is suppressed at an initial rate of $\approx - 12$ K/kbar, with no clear feature of the transition being visible for $P > 3.5$ kbar.  The high pressure, drop in resistivity transition temperature increases with increasing pressure at a rate of  $\approx 17$ K/kbar for $P > 5.5$ kbar.  It is worth noting that neither phase transition is seen in the 5.1 kbar data set, but also note that given the opposite sign of the resistive anomalies, this may be a coincidence.  In addition, it is also worth noting that there is no evidence of a crossing, of these two transitions, i.e. there is no evidence, in the transport data, that either phase transition exists as a lower phase transition, below a higher one.

In precisely the pressure region where the two high temperature phase transitions are either dropping to zero in a very non-linear fashion or fading out / into each other, superconductivity can be detected at low temperature.  The sharpest transitions are found for 3.5 and 5.1 kbar.  The transitions seen in the 2.3 kbar and 8.6 kbar data have long, low temperature "feet" that extend to below our 2 K minimum temperature and are shown with an appropriate error bars in Fig. \ref{F3}.  The position of the superconducting dome relative to the two higher temperature phase transitions is even more clearly illustrated when the low temperature resistivity as a function of pressure is examined in conjunction with the $P - T$ phase diagram in Fig. \ref{F3}.  The resistivity for $T = 15$ K was chosen since it is well above the superconducting transitions (as well as the broad down-turns) and yet is in a fairly temperature independent, low temperature residual resistivity region.  For low pressures, $P \le 3.5$ kbar, there is a gradual reduction of low temperature scattering as pressure is increased.  For 3.5 kbar $< P <~8.6$ kbar there is a sharp decrease in the low temperature scattering as pressure is increased, resulting in an over five times reduction in the low temperature resistivity between $P = 3.5$ kbar and 8.6 kbar.  For $P > 8.6$ kbar the low temperature resistivity is essentially pressure independent and low.  The superconducting dome exists in precisely the pressure range that is associated with this dramatic decrease in low temperature scattering (see inset to Fig. \ref{F3}).  To summarize:  pressure induced superconductivity in the CaFe$_2$As$_2$ system only appears when (i) the structural (antiferromagnetic) phase transition has been dramatically reduced or fully suppressed, and (ii) when the higher pressure, loss of scattering phase transition has not been too fully established at low temperatures, e.g the reduction of resistivity associated with the higher pressure transition has not been completely realized.

Clearly the nature of the higher temperature phase transitions will be key to understanding the mechanism for the superconductivity in CaFe$_2$As$_2$, and by analogy, for all of the FeAs based superconductors.  The low pressure phase transition is reasonably identified as a tetragonal to orthorhombic (and antiferromagnetic), structural phase transition. The higher pressure transition, with its drop in resistivity, on the other hand, could have many origins: electronic, magnetic and/or structural. It appears though that going too far into this new phase is detrimental to superconductivity. With this in mind, then, one possible interpretation of data presented in Fig. \ref{F3} is that superconductivity exists when the structural (magnetic) phase transition is suppressed and there are still sufficient fluctuations or excitations to allow for the coupling / interactions required to allow Cooper pair formation.  If these fluctuations or excitations are too fully removed, as manifest by too great of a reduction of the low temperature resistivity, then superconductivity does not persist.

It is not clear yet what precisely happens to the two, high temperature phase transitions in the vicinity of 5 kbar.  It is possible that both transitions drop rapidly to zero, in which case the P-T phase diagram shown in Fig. \ref{F3} bears a striking resemblance to a pressure induced quantum critical point.  The superconducting dome being centered around the critical pressure is also consistent with this picture.  Further data, perhaps from other measurement techniques, will be needed to clarify this key region of the phase diagram.

Speculation aside, it is manifestly clear that further study of CaFe$_2$As$_2$ under pressure holds great promise to unraveling the puzzles associated with superconductivity in FeAs compounds.  Very modest changes in hydrostatic pressure move this compound from manifesting the tetragonal to orthorhombic phase transition ubiquitous to the AFe$_2$As$_2$ (A = Ca, Sr, Ba) compounds, to manifesting superconductivity below 12 K, to stabilizing a phase transition that increases rapidly with pressure and manifests a dramatic reduction of low temperature resistivity.  Studies of either NMR or neutron scattering under pressure should bring new insights to the high temperature transitions seen in this very promising compound.

\begin{acknowledgments}
Work at the Ames Laboratory (sample growth, characterization and data analysis)) was supported by the US Department of Energy - Basic Energy Sciences under Contract No. DE-AC02-07CH11358.  MST (hydrostatic pressure resistance measurements and data analysis) gratefully acknowledges support of the National Science Foundation under DMR-0306165 and DMR-0805335.

\end{acknowledgments}

\clearpage

\clearpage

\clearpage

\begin{thebibliography}{99}

\bibitem{1}  Y. Kamihara, T. Watanabe, M. Hirano, H. Hosono, Journal of the American Chemical Society  {\bf 130}, 3296 (2008).

\bibitem{2}  H. Kito, H. Eisaki, A. Iyo, J. Phys. Soc. Jpn. {\bf 77}, 063707 (2008).

\bibitem{3}  M. Rotter, M. Tegel, D. Johrendt, arXiv:0805.4630, unpublished.

\bibitem{4}  G. F. Chen, Z. Li, G. Li, W. Z. Hu, J. Dong, X. D. Zhang, P. Zheng, N. L. Wang, and J. L. Luo, arXiv:0806.1209, unpublished.

\bibitem{ger} M. Pfisterer, G. Nagorsen, Z. Naturforschung B {\bf 53}, 703 (1980). 

\bibitem{5}  N. Ni, S. Nandi, A. Kreyssig, A. I. Goldman, E. D. Mun, S. L. Bud'ko, P. C. Canfield, arXiv:0806.4328, unpublished.

\bibitem{6}  G. Wu, H. Chen, T. Wu, Y. L. Xie, Y. J. Yan, R. H. Liu, X. F. Wang, J. J. Ying, and X. H. Chen, arXiv:0806.4279, unpublished.

\bibitem{7}  F. Ronning, T. Klimczuk, E.D. Bauer, H. Volz, J.D. Thompson, J. Phys.: Condens. Matter, {\bf 20} 322201 (2008).

\bibitem{ngol} A. I. Goldman, D. N. Argyriou, B. Ouladdiaf, T. Chatterji, A. Kreyssig, S. Nandi, N. Ni, S. L. Bud'ko, P.C. Canfield, R. J. McQueeney, arXiv:0807.1525.

\bibitem{8}  P. C. Canfield, Z.  Fisk, Phil. Mag. B  {\bf 65}, 1117 (1992).

\bibitem{eil81a} A. Eiling, J. S. Schilling, J. Phys. F: Met. Phys. {\bf 11}, 623 (1981).


\end{thebibliography}
\end{document}